\documentclass[12pt]{iopart}

\newcommand{\unit}[1]{\ensuremath{\;\mathrm{#1}}}

\usepackage{graphicx}
\usepackage{hyperref} 
\hypersetup{bookmarks=true,colorlinks=true,linkcolor=blue,citecolor=green}

\begin{document}


\title{Studying Free-Space Transmission Statistics and Improving Free-Space QKD in the Turbulent Atmosphere}


\author{C. Erven,$^1$ B. Heim,$^{1,2,3}$ E. Meyer-Scott,$^1$ J.P. Bourgoin,$^1$ R. Laflamme,$^{1,4}$ G. Weihs$^{1,5}$ and T. Jennewein$^{1}$}
\address{$^1$ Institute for Quantum Computing, Department of Physics and Astronomy, University of Waterloo, 200 University Avenue West, Waterloo, ON, N2L 3G1, Canada}
\address{$^2$ Max Planck Institute for the Science of Light, G\"unther-Scharowsky-Str. 1, Building 24, 91058 Erlangen, Germany}
\address{$^3$ Erlangen Graduate School in Advanced Optical Technologies (SAOT), University of Erlangen-Nuremberg, Paul-Gordan-Str. 6, 91052 Erlangen, Germany}
\address{$^4$ Perimeter Institute, 31 Caroline Street North, Waterloo, ON, N2L 2Y5, Canada}
\address{$^5$ Institut f\"ur Experimentalphysik, Universit\"at Innsbruck, Technikerstrasse 25, 6020 Innsbruck, Austria}
\eads{\mailto{cerven@iqc.ca}, \mailto{bettina.heim@mpl.mpg.de}, \mailto{thomas.jennewein@uwaterloo.ca}}

\date{\today}

\begin{abstract}
The statistical fluctuations in free-space links in the turbulent atmosphere are important for the distribution of quantum signals. To that end, we first study statistics generated by the turbulent atmosphere in an entanglement based free-space quantum key distribution (QKD) system. Using the insights gained from this analysis, we study the effect of link fluctuations on the security and key generation rate of decoy state QKD concluding that it has minimal effect in the typical operating regimes. We then investigate the novel idea of using these turbulent fluctuations to our advantage in QKD experiments. We implement a signal-to-noise ratio filter (SNRF) in our QKD system which rejects measurements during periods of low transmission efficiency, where the measured quantum bit error rate (QBER) is temporarily elevated. Using this, we increase the total secret key generated by the system from 78,009\unit{bits} to 97,678\unit{bits}, representing an increase of 25.2\% in the final secure key rate, generated from the \emph{same} raw signals. Lastly, we present simulations of a QKD exchange with an orbiting LEO satellite and show that an SNRF will be extremely useful in such a situation, allowing many more passes to extract a secret key than would otherwise be possible.
\end{abstract}

\maketitle


\section{Introduction}

Quantum key distribution (QKD), one of the first experimentally realizable technologies from the field of quantum information, has by now seen a number of robust implementations both in fibre \cite{SFIK11,PPAB09,SBCDLP09,GRTZ02} and free-space \cite{JRYYZXWYHJYYCPP10,EBHWSL09,ECLW08,PFHLK08,UTSWSLBJPTOFMRSBWZ07,NHMPW02}. Indeed, it has already reached the level of maturity so as to be offered as a commercial product from a number of companies \cite{MagiQ,idQuantique,QuintessenceLabs,SeQureNet}. While the fastest systems to date are based on fibre transmission media \cite{DYDSS08}, they will remain limited to a transmission distance of about 200\unit{km} until reliable quantum repeaters are realized. Even taking into account expected future advances in fibre, source, and detector technology, secure key distribution will still be limited to about 400\unit{km} using fibres.

QKD with orbiting satellites has long been proposed as a solution for global key distribution, as evidenced by the growing number of feasibility studies that have been conducted \cite{NHMPW02,RTGK02,AJPLZ03,AFMMCPAJUSBRLBWZ07,BMHHEHKHDGLLJ12}. QKD with low earth orbit (LEO) satellites likely represents the most feasible solution since they will have the shortest free-space transmission distance with the lowest losses. However, LEO satellites travel quickly with short orbital periods limiting the time available to perform QKD during a single pass to the order of 300\unit{sec} \cite{BMHHEHKHDGLLJ12,NHMPW02}. Thus, it is important to have a thorough understanding of the transmission properties of the free-space channel which the photons will travel through in order to properly evaluate the feasibility of such a system. As well, with such a short time to exchange a key, it is important to extract the most secure key bits from the relatively small number of signals sent and received during a pass.

To these ends, this article first examines some recent theoretical work on the transfer of quantum light and entanglement through the turbulent atmosphere; then experimentally determined free-space transmission efficiency curves measured with an entanglement based free-space QKD system are analyzed; this is followed by a discussion of the implications of link fluctuations on decoy state QKD; a method for improving free-space QKD key rates in the turbulent atmosphere through the use of a signal-to-noise ratio filter (SNRF) is then put forward; followed by the experimental results of implementing such a filter and their implications for the security of the system.

\section{Free-Space Optical Link Statistics}

The propagation of classical light through turbulent atmosphere has long been of interest in theoretical investigations, including such diverse phenomena as diffraction, scintillation, and the absorption of light by molecules in the atmosphere which produce beam wander and broadening \cite{Kol41,Tat71,Fan75,Fan80,AP05,APY95}. Satellite based communication has also been investigated in the context of a turbulent atmosphere \cite{Fri67,Sha11,AP05,APY95}. From these studies it has been shown that the intensity fluctuations due to the turbulent atmosphere can be assumed to be log-normally distributed in the regime of weak fluctuations and strong losses. This has also been confirmed in various experiments (see e.g. \cite{MCPH04}).

Recently, Vasylyev, Semenov and Vogel \cite{VSV12,SV10,SV09} have provided a theoretical foundation for studying the influence of fluctuating loss channels on the transmission of quantum and entangled states of light. Like others \cite{Smi93,MCPH04}, in Refs. \cite{SV10,SV09} they approximate the probability distribution of the (fluctuating) atmospheric transmission coefficient (PDTC) in the case of entanglement distribution according to the log-normal distribution:
\begin{equation}\label{eq.PDTC}
  \mathcal{P}(\eta_{atm}) = \frac{1}{\sqrt{2 \pi} \sigma \eta_{atm}} \exp \bigg[- \frac{1}{2} \bigg(\frac{\ln \eta_{atm} + \bar{\theta}}{\sigma} \bigg)^{2} \bigg]
\end{equation}
where $\eta_{atm}$ is the atmospheric transmittance, $\overline{\theta} = -\ln<\eta_{atm}>$ is the logarithm of the mean atmospheric transmittance, and $\sigma$ is the variance of $\theta = -\ln \eta_{atm}$ characterizing the atmospheric turbulence.

Equation \ref{eq.PDTC} only describes in a simplified way the transmission property of an atmospheric channel and ignores any phase (front) fluctuations. This is sufficient for our analysis because our experiments utilize the direct detection of single photons, making the phase nature of the transmission irrelevant.

\subsection{Measuring Free-Space Link Statistics with Entangled Photons}

To begin, we measured the free-space transmission efficiency statistics in our entanglement based QKD system. The system is comprised of a compact Sagnac interferometric entangled photon source \cite{EHRLW10,KFW06,FHPJZ07} operating at 808\unit{nm}, a 1,305\unit{m} free-space optical link where the outgoing/incoming beam is expanded/contracted by the use of appropriate telescopes (the telescopes have a 75\unit{mm} collection lens and a 25:1 magnification), two compact passive polarization analysis modules, quad module silicon avalanche photodiode single photon detectors (PerkinElmer SPCM-AQ4C, $\sim$50\% detection efficiency, $\sim$400 dark counts/sec), time-tagging units, GPS time receivers, two laptop computers, and custom written software \cite{ECLW08}. We choose to work at a wavelength of 808\unit{nm} to take advantage of a peak in the typical atmospheric transmission \cite{BMHHEHKHDGLLJ12} as well as high detection efficiency at that wavelength in our detectors. Usually there is a 10\unit{nm} interference filter at the entrance of the polarization detector box used to reject background light; however, we remove it for all experiments in this paper in order to simulate a scenario (such as a satellite link) with a higher background noise level in order to test the usefulness of our signal-to-noise ratio filter, described later.

Brida \emph{et al.} \cite{BGN00} were the first to suggest using two photon entangled states for the absolute quantum efficiency calibration of photodetectors. We adapt their method here to measure the PDTC of the free-space channel by first performing a local experiment with the same equipment (source, polarization analyzers, photon detectors) so that we can measure the various other efficiencies of the system. Then through comparison of the experiments performed locally and over the free-space we can extract the PDTC of the link.

In a local experiment we expect the number of counts per second seen by Alice ($N_{A}$) and Bob ($N_{B}$) to be given by
\begin{equation}\label{eq.AliceLocalSingles}
  N_{A} = N \eta_{A} = N \eta_{A_{source}} \eta_{A_{pol}} \eta_{A_{det}}
\end{equation}
\begin{equation}\label{eq.BobLocalSingles}
  N_{B} = N \eta_{B} = N \eta_{B_{source}} \eta_{B_{pol}} \eta_{B_{det}}
\end{equation}
where $N$ is the total number of pairs produced at the source per second, $\eta_{A}$ is Alice's total transmission efficiency (comprised of the source coupling efficiency, $\eta_{A_{source}}$, polarization analyzer efficiency, $\eta_{A_{pol}}$, and detector efficiency, $\eta_{A_{det}}$), and similarly for Bob. Additionally, the expected number of observed coincidences per second ($N_{coin}$) between Alice and Bob, found using a coincidence window ($\Delta t_{coin}$) to identify entangled photon pairs, is given by
\begin{equation}\label{eq.LocalCoins}
  N_{coin} = N \eta_{A} \eta_{B}.
\end{equation}
Dividing the measured coincidence count rate ($N_{coin}$) by the observed singles rate at Alice ($N_{A}$) yields an estimate for the total loss caused by Bob's optics ($\eta_{B}$) including the source coupling, polarization analyzer, and photon detectors. Double pair emissions, where two photon pairs are created in the source crystal at once, could lead to corrections in Eqs. \ref{eq.LocalCoins}-\ref{eq.LinkAccidentals} at sufficiently high pump powers. However, for the experiments detailed here, the pumping strength was sufficiently low that double pair emissions were negligible and thus safely ignored.

For experiments performed over the free-space link, the equation for Bob's singles rate gets modified to
\begin{eqnarray}\label{eq.BobLinkSingles}
  N_{B} & = & N \eta_{B} + N_{background} \nonumber \\
   & = & N \eta_{B_{source}} \eta_{B_{atm}} \eta_{B_{pol}} \eta_{B_{det}} + N_{background}
\end{eqnarray}
where his total transmission efficiency, $\eta_{B}$, now includes a term for the link transmission efficiency, $\eta_{B_{atm}}$, and an additional term, $N_{background}$, is added representing background photons which are collected and measured by Bob's receiver. The equation for the coincidence rate is similarly modified to
\begin{equation}\label{eq.LinkCoins}
  N_{coin} = N \eta_{A} \eta_{B} + N_{accidental}
\end{equation}
where $N_{accidental}$ represents accidental coincidences of Alice's measurements with the background photons measured by Bob. Fortunately, the accidental rate given to good approximation by
\begin{equation}\label{eq.LinkAccidentals}
  N_{accidental} \approx N_{A} N_{B} \Delta t_{coin}
\end{equation}
can be easily estimated by finding the number of coincidences between Alice's measurements and Bob's measurements shifted by a few coincidence windows and then subtracted from the results.

To find the free-space link PDTC we divide the coincidence rate ($N_{coin}$) observed during a link experiment by Alice's local single photon count rates ($N_{A}$) which gives the PDTC for Bob's total loss, $\eta_{B}$, including all of the losses in his equipment. Then, using the estimate from the local experiment, we divide out the losses from Bob's equipment leaving only the atmospheric transmission, $\eta_{B_{atm}}$, allowing us to construct the PDTC for the free-space channel. There is an alternative method for estimating the free-space link PDTC using only the singles rates from an experiment over a free-space link. However, the method just described using coincidences is more accurate than using just the singles rates since the only source of error is the accidental coincidence rate ($N_{accidental}$) which we can estimate and remove.

\begin{figure}[hbt]
    \centering
    \includegraphics[width=0.8\textwidth]{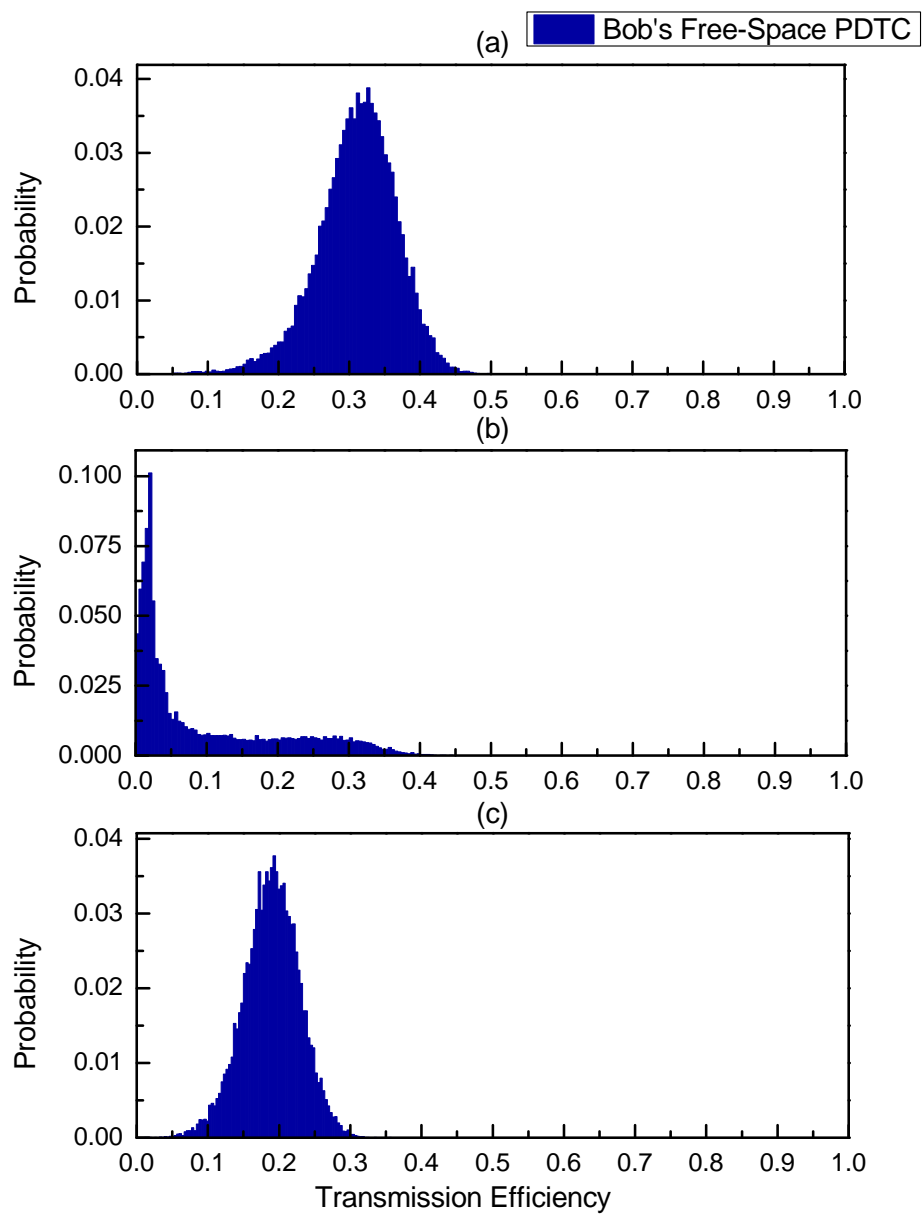}
    \caption{Probability distribution of the transmission coefficient (PDTC) for the case of (a) an optimized free-space channel, (b) a free-space channel with artificially increased turbulence using a heat gun placed in front of the sending telescope, and (c) a free-space channel where the beam is defocused in order to simulate larger losses. The detection (sampling) time was 10\unit{ms}.}
    \label{fig.IQCPDTC}
\end{figure}

We studied three different scenarios with our system for the distribution of entangled photons over free-space channels corresponding to the following conditions: a maximum free-space transmission with optimized pointing and focusing parameters (Fig. \ref{fig.IQCPDTC} (a)), a transmission with artificially increased turbulence using a heat gun to heat the air immediately in front of the sending telescope (Fig. \ref{fig.IQCPDTC} (b)), and a defocused transmission as a way to simulate larger losses (Fig. \ref{fig.IQCPDTC} (c)). For each of these experiments, the data was broken up into blocks of a certain duration which we call the block duration and then the efficiency was estimated for each block using the method described above. These results are then summed up into a histogram, normalized, and displayed as the PDTC for that link. In all cases, the distributions are shown with a block duration of 10\unit{ms} since it has been shown that this is the typical timescale for atmospheric turbulence\cite{MCPH04}. All measurements were performed on August 24, 2011 between the hours of 12 and 1am since the system requires the reduced background experienced at night to operate, with a total data acquisition time for each experiment of 3 minutes. The temperature was approximately 18$^{\circ}$C with clear visibility in an urban environment with typical city illumination levels.

Fig. \ref{fig.IQCPDTC} (a) shows that we experienced extremely good atmospheric conditions during the experiments since the observed transmission coefficient for the well aligned link was very close to a Poissonian distribution. The term Poissonian here really refers to the original graphs of integer photon counts versus the frequency with which they were observed. We would expect the transmission coefficient for a local system without a free-space link to be Poissonian in nature owing to the pair creation process and detection. The fact that we still observe a Poissonian distribution with a free-space link implies that our atmospheric conditions were very good since the presence of the link did not alter the nature of the statistics.

The defocussed transmission case, Fig. \ref{fig.IQCPDTC} (c), is also very close to a Poissonian distribution only narrowed with a decreased overall transmittance compared to Fig. \ref{fig.IQCPDTC} (a). This is expected since defocusing the beam increased it to a size larger than the receiver telescope thus causing fluctuations in the transmission efficiency experienced over the free-space link to be smoothed out (ie. causing it to be even closer to a Poissonian distribution) while at the same time lowering the overall transmittance since many more photons missed the receiver telescope and consequently were not collected and measured. For the experiment where turbulence was artificially added by letting the beam pass over hot air produced by a heat gun, Fig. \ref{fig.IQCPDTC} (b), the distribution indeed changes towards a log-normal distribution as predicted.

\section{Effect of Link Fluctuations on Decoy State QKD}

Having investigated the PDTC for a number of different free-space channels in the previous section, we now turn our attention to the question of what effect atmospheric turbulence might have on weak coherent pulse QKD with decoy states. Attenuated lasers, while convenient for QKD, do not emit true single photons but rather a mixture of photon number states following a Poissonian distribution. This limits the distance over which QKD can be performed as Eve can perform a photon number splitting attack to gain full information on multi-photon pulses \cite{GLLP04}. This attack relies on Eve's ability to block single photon pulses and thus modifies the channel transmission nonlinearly depending on the photon number. However, this attack can be detected through the use of decoy states of various pulse strengths \cite{Hwa03,LMC05}, and an additional step in the security phase which verifies that the channel transmission does not depend on the mean photon number. Thus, it is crucial for free space QKD systems using decoy states to consider atmospheric fluctuation since the security of the protocol depends strongly on the relative transmission of the various pulse strengths.

\begin{figure}[htp]
    \centering
    \includegraphics[width=0.9\textwidth]{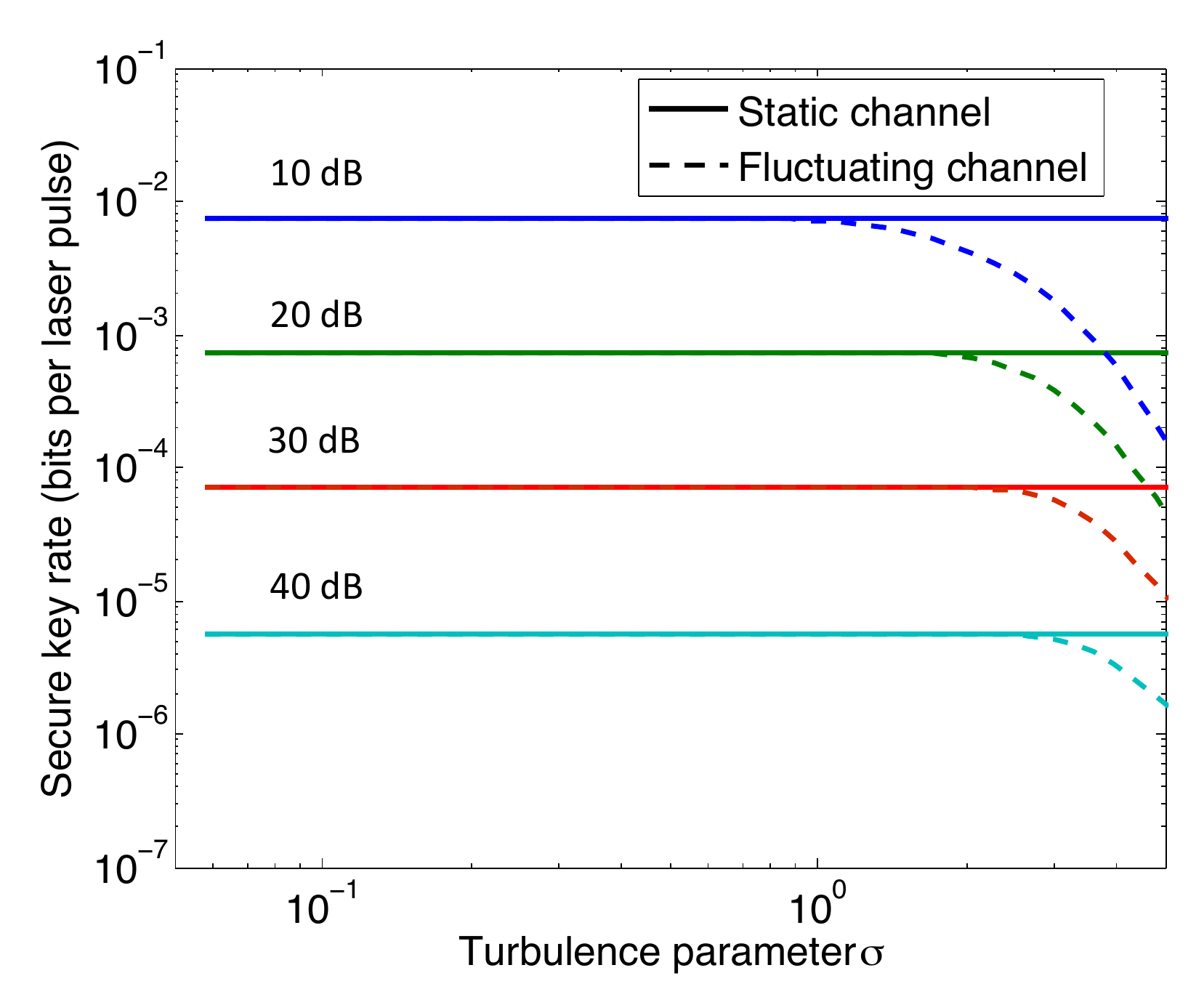}
    \caption{Secure key rate versus the turbulence parameter, $\sigma$, comparing static (solid line) and fluctuating channel (dashed line) with same mean loss. Average channel losses are indicated for the four curves. Deviation is only apparent at very strong turbulence, meaning the static channel approximation is sufficient for most cases.}
    \label{fig.WCPvsSigma}
\end{figure}

\begin{figure}[htp]
    \centering
    \includegraphics[width=0.9\textwidth]{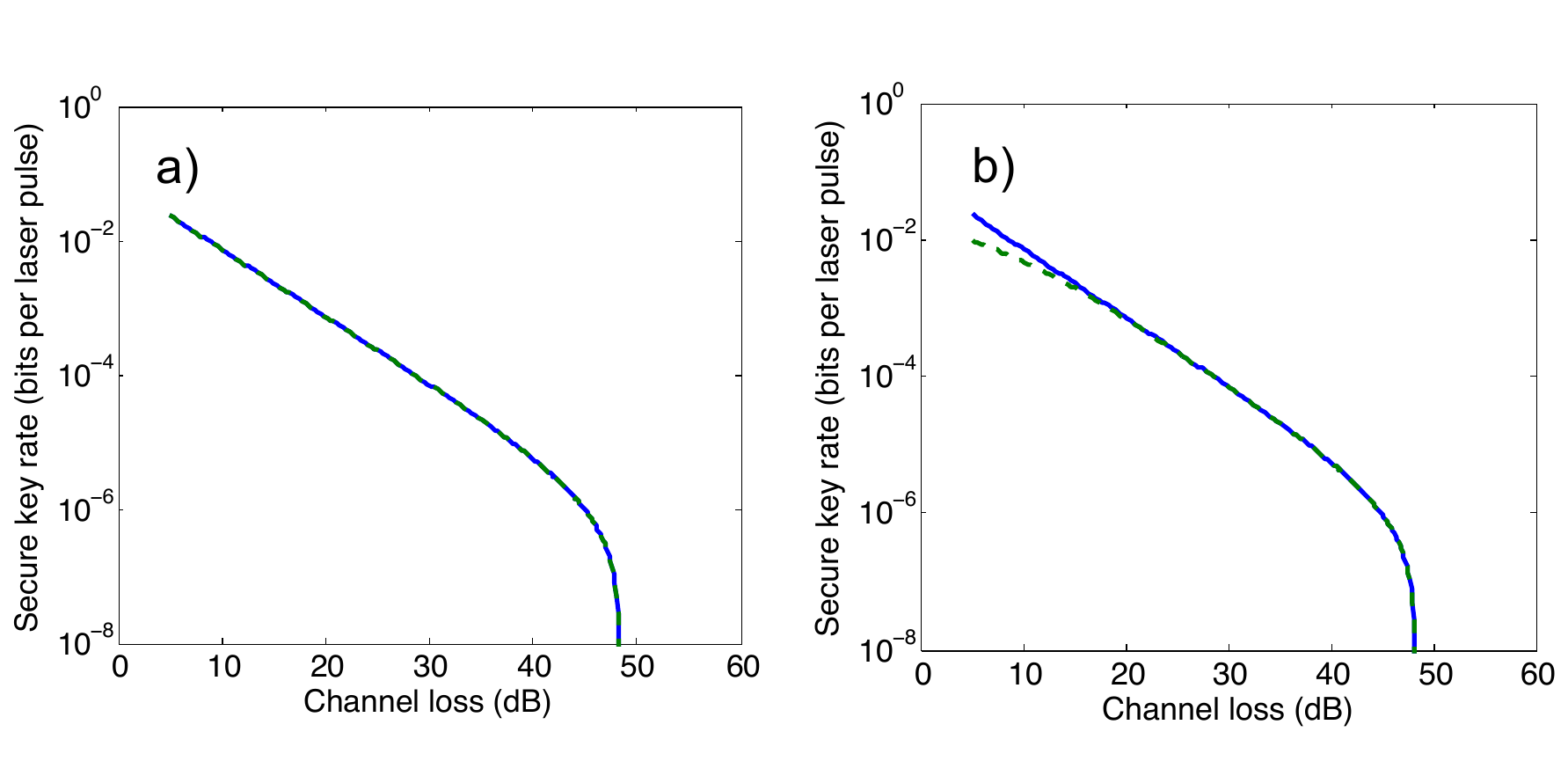}
    \caption{Secure key rate versus loss, comparing a static quantum channel (solid line) to a fluctuating free-space quantum channel (dashed line) with same mean loss. Figure a) considers ``good'' atmosphere with $\sigma=0.18$, resulting in no deviation between the static and fluctuating channel. Figure b) considers ``bad'' atmosphere with $\sigma=1.8$, resulting in less secure key for the fluctuating channel at low loss.}
    \label{fig.WCPvsLoss}
\end{figure}

Here we investigate whether the assumption of a static channel for determining secure key length is valid when the channel is, in reality, fluctuating. We consider a one-decoy protocol from Ma \emph{et al.} \cite{MQZL05}, including the ``tighter bound'' from section E.2, along with the PDTC generated from the photon statistics in atmosphere taken from \cite{MCPH04}, and with a realistic error correction efficiency of $f(e) = 1.22$ used. Figures \ref{fig.WCPvsSigma} and \ref{fig.WCPvsLoss} compare the results from a simulation of secure key rates based on a simple static channel versus a channel fluctuating with a log-normal distribution.

Fig. \ref{fig.WCPvsSigma} shows that approximating a fluctuating channel as a static channel with the same mean loss is sufficient so long as the atmosphere is not extremely turbulent. It should be noted that the turbulence model of Milonni \emph{et al.} \cite{MCPH04} begins to fail at extremely high turbulence levels possibly leading to inaccuracies in the extreme right portion of Fig. \ref{fig.WCPvsSigma}. Unfortunately, we are unaware of a correct high turbulence free-space link model to replace it with in this regime. Nevertheless, we can still draw important conclusions in the usual operating scenarios. Further, at moderate turbulence strengths Fig. \ref{fig.WCPvsLoss} shows that the static channel approximation is valid as long as the channel loss is above $\sim$15\unit{dB}, a typical condition in long distance free-space QKD (see Ref.~\cite{BMHHEHKHDGLLJ12} for example scenarios). Therefore, the security of weak coherent pulse QKD with decoy states is not significantly affected by a fluctuating free-space quantum channel as compared to the usual assumed static channel since the differences only arise in a situation where the high turbulence would likely make a successful transmission impossible or with a link with such low losses that the transmission distance is likely uninteresting. This also paves the way for checking whether the key rates could possibly be improved with a signal-to-noise ratio filter.

\section{Improving QKD with a Signal-to-Noise Ratio Filter}

Using the link statistics analysis and the data from the experiments above, we now investigate the use of a signal-to-noise ratio filter (SNRF) in order to increase the final key rate in QKD systems with a turbulent quantum transmission channel. The idea of the SNRF is to throw away data blocks where the signal-to-noise ratio (SNR) was low based on a directly measurable quantity, the signal strength, under the assumption that the noise caused by background events remains constant. While this has the consequence of decreasing the overall raw key rate, it is possible to actually improve the final secret key rate since we omit the blocks where the SNR was lower and correspondingly the QBER was inflated by the larger relative contribution from the background. We mention that similar filtering techniques have been explored in the continuous variable regime, though the effect of fading on CV quantum states is very different. Fading and excess noise quickly destroy Gaussian quantum features such as entanglement and purity of states; however, different possibilities have been offered to recover these \cite{HMDFLLA06,DLHMFLA08}.

We define the SNRF algorithm as follows. One begins by measuring the background contribution of the quantum free-space channel (in terms of a count rate) with the entangled source switched off. Then one defines the singles contribution from the source divided by the background contributions as the dimensionless SNR. One then throws away low SNR blocks where the background contribution is proportionally higher according to a preset SNR threshold. The idea can also be mapped to real coincidences from the source divided by background coincidences where these numbers now implicitly depend on the coincidence window used.

In the following, we implement the equivalent algorithm where rather than using the dimensionless SNR we instead use the singles rates to define our threshold. The SNR threshold is implicitly used in this protocol since the background noise is assumed to remain constant. Thus, examining the optimum singles rate threshold effectively amounts to finding the optimum SNR threshold since one could calculate this number by first measuring the background, subtract it from the total measured singles, and then divide the remainder by the measured background to arrive at the SNR. In the remainder of this paper we will refer to all such equivalent protocols as a SNRF algorithm.

\begin{figure}[htp]
    \centering
    \includegraphics[width=0.8\textwidth]{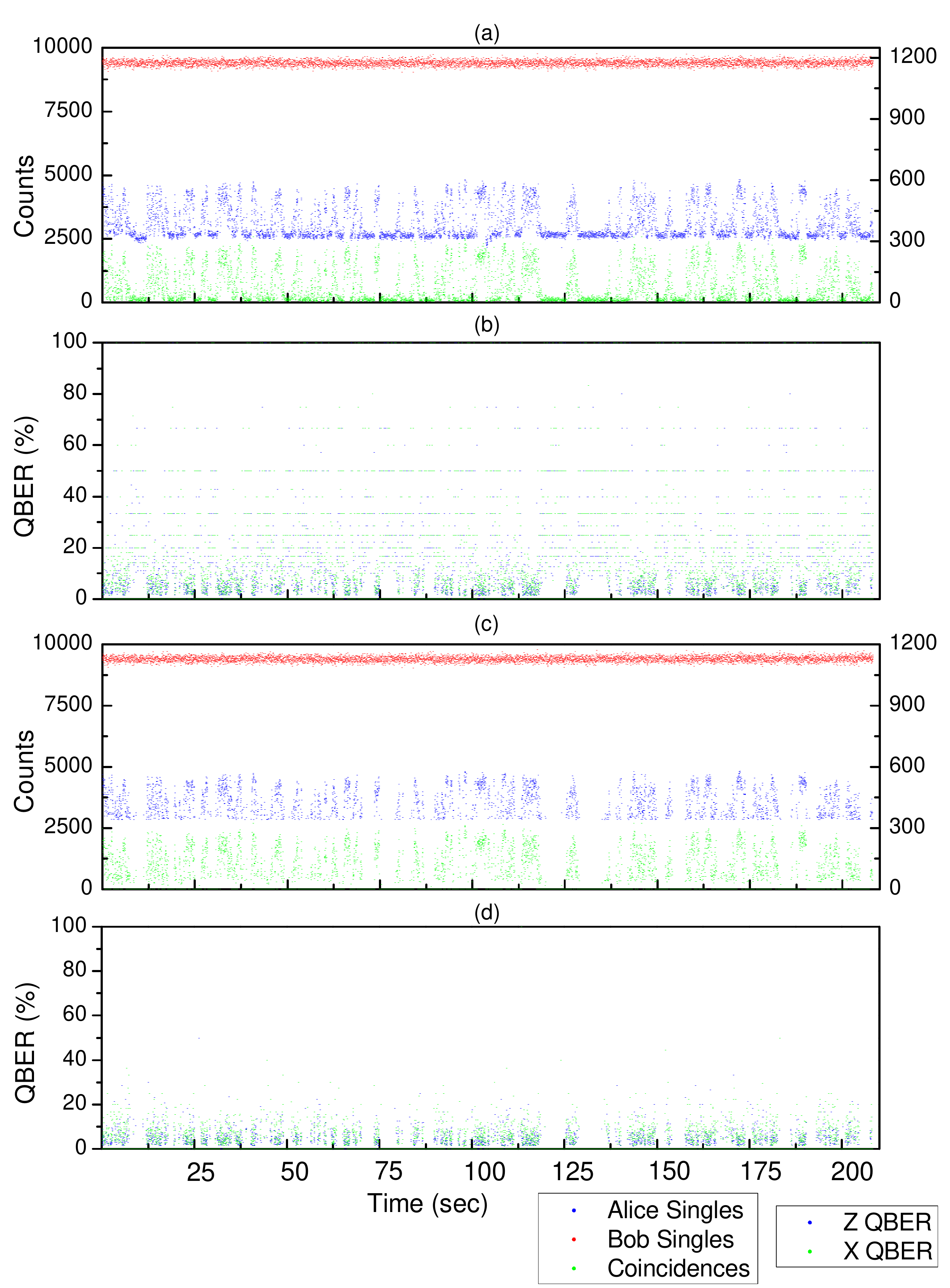}
    \caption{Alice's local single count rate (red curves, left axis), Bob's single count rate measured over the link (blue curves, left axis), the coincidence rate (green curves, right axis), and QBER in the Z (blue curve) and X (green curve) bases for the high free-space turbulence experiment of Fig. \ref{fig.IQCPDTC} (b) for the case of (a-b) no SNRF and (c-d) the optimum SNRF of 95,000\unit{singles/sec}. The data points are grouped according to the optimum block duration of 30\unit{ms} and a coincidence window of 5\unit{ns} is used.}
    \label{fig.SingleCoinQBER}
\end{figure}

Fig. \ref{fig.SingleCoinQBER} shows (a) Alice's local rates (red curve) and Bob's singles count rates measured over the link (blue curve) along with the coincidence count rate (green curve) and (b) the corresponding QBER's measured in the Z (blue curve) and X (green curve)\footnote{The Z basis here refers to the basis of the Pauli $\sigma_{Z}$ operator (ie. horizontal and vertical polarization), while the X basis refers to the basis of the Pauli $\sigma_{X}$ operator (ie. +45$^{\circ}$ and -45$^{\circ}$ polarization).} bases when no SNRF is used for the artificially increased turbulence experiment of Fig. \ref{fig.IQCPDTC} (b). Whereas, Fig. \ref{fig.SingleCoinQBER} (c) shows Alice and Bob's singles and coincidence rates when the optimum SNRF threshold of 95,000\unit{counts/sec} (discussed below) is applied, and (d) shows the corresponding QBER. The data points are grouped according to the optimum block duration of 30\unit{ms} (thus, each data point represents 30\unit{ms} worth of data) and a coincidence window of 5\unit{ns} is used. Here one can clearly see the high background detection rate experienced by Bob (a situation that will be typical of a QKD link performed to an orbiting satellite) as the flat bottom of his singles rate graph (Fig. \ref{fig.SingleCoinQBER} (a) blue curve), as well as the wildly varying coincidence rates (Fig. \ref{fig.SingleCoinQBER} (a) green curve) where the points close to the x-axis largely consist of accidental coincidences. From Fig. \ref{fig.SingleCoinQBER} (a) blue curve, we can estimate Bob's mean background count rate at roughly 2,700\unit{counts/sec}. Each background count which is registered as a valid coincidence will be uncorrelated with Alice's measurements and contribute a 50\% error rate to the QBER.

The SNRF idea is neatly illustrated here by looking at the many high QBER values, corresponding to the low signal phases in Fig. \ref{fig.SingleCoinQBER} (b) associated with Bob's low singles and coincidence rates from the top graph. We know from the experiment corresponding to Fig. \ref{fig.IQCPDTC} (a) for the well aligned link that the intrinsic QBER is {$\sim$2.34\%}; however; the QBER observed for the turbulent link corresponding to Fig. \ref{fig.IQCPDTC} (b) and Fig. \ref{fig.SingleCoinQBER} (a) and (b) was instead {$\sim$5.51\%}. This increase in the measured QBER over the actual QBER of the system will lower the final secret key rates. However, one can see that when the low SNR regions are removed from the singles and coincidence graph (Fig. \ref{fig.SingleCoinQBER} (c)) using the optimum SNRF, many of the corresponding high QBER blocks (Fig. \ref{fig.SingleCoinQBER} (d)) are also removed. Thus, we are able to lower the measured QBER from $\sim$5.51\% to $\sim$4.30\%, a value closer to the intrinsic error rate of the system, allowing the system to generate many more secret key bits than would otherwise be possible.

The secret key rate formula for our system expressed in secret key bits per raw key bit is given by \cite{MFL07b}
\begin{equation}\label{eq.InfiniteKeyRate}
  R = \frac{1}{2}(1 - f(e)h_{2}(e) - h_{2}(e))
\end{equation}
where $f(e)$ is the error correction inefficiency as a function of the error rate, normally $f(e) \geq 1$ with $f(x) = 1$ at the Shannon limit, and $h_{2}(e) = -e \log e - (1-e) \log (1-e)$ is the binary entropy function. For the clarity of the argument we have used the infinite key limit formula; however, the insights gained should transfer to the finite key limit. Looking at Eq. \ref{eq.InfiniteKeyRate}, we can see that a higher QBER is detrimental to the final key rate for two reasons (a)~increased error correction inefficiency and (b)~increased privacy amplification. The Cascade algorithm \cite{BS94,SY00} and low density parity check (LDPC) codes \cite{Gal62,MN97,Pea04,MCMHT09} are the two most commonly employed error correction algorithms used in QKD systems. As the QBER climbs the number of parities revealed (and correspondingly the information about the key which has to be accounted for in privacy amplification) increases. This applies even in the ideal case of error correction algorithms operating at the Shannon limit. Privacy amplification is then used after error correction to squeeze out any potential eavesdropper and ensure that the probability that anyone besides Alice and Bob knows the final key is exponentially small at the cost of shrinking the size of the final key. Privacy amplification is commonly accomplished by applying a two-universal hash function \cite{CW79,Kra94} to the error corrected key and then using Eq. \ref{eq.InfiniteKeyRate} to determine how many bits from this operation may be kept for the final secret key. Both the number of bits exposed during error correction and the measured QBER are used to determine the final size of the key. Additionally, the secure key rate formula is a non-linear function of the QBER so that decreasing the QBER does better than a linear improvement in the final key rate \cite{CCHGKMTONMSWFRGGHTB11}. Thus, the fewer parities revealed during error correction and the lower we can make the measured QBER, the larger the final key will be.

The use of a SNRF could potentially open a loophole in the security proofs for QKD since we are now discarding data (which is typically not allowed by the proofs) depending on Bob's measured singles rates. However, we are implementing the SNRF on Bob's singles rate which is a sum over all of his detectors during a block of data, so the SNRF is detector independent. Additionally, discarding data should be equivalent to a decrease in the channel transmission efficiency (which could happen anyways due to atmospheric effects) and thus should not affect the security proof. One might initially think that measuring the SNRF in real-time could thwart a potential eavesdropper; however, any real-time measurement would require a probe beam either of a different wavelength or propagating in a different spatial mode so that it could be easily separated from the signal photons. This difference would easily allow an eavesdropper to fake whatever background they wished on the probe beam while leaving the signal beam untouched. Therefore, for this paper we assume that using a SNRF does not compromise the security of our system; however, we stress that it remains an open question whether security can be proven for this scenario. We also point out that for an entangled QKD protocol security does not depend on the transmission of the quantum channel; whereas, if one wanted to use a SNRF in a decoy state protocol, which works by measuring the channel gain for each photon number component, the issue of security would be delicate and require careful analysis so as not to open up any security loopholes. We hope that by showing the utility of the SNRF methods we can stimulate work on proving its security. Finally, we also mention recent work by Usenko \emph{et al.} \cite{UHPWMLF12} which theoretically studied the influence of fading on Gaussian states in the framework of the continuous variable QKD filtering ideas mentioned earlier.

\section{Experimental Results and Discussion}

\begin{figure}[htp]
    \centering
    \includegraphics[width=0.9\textwidth]{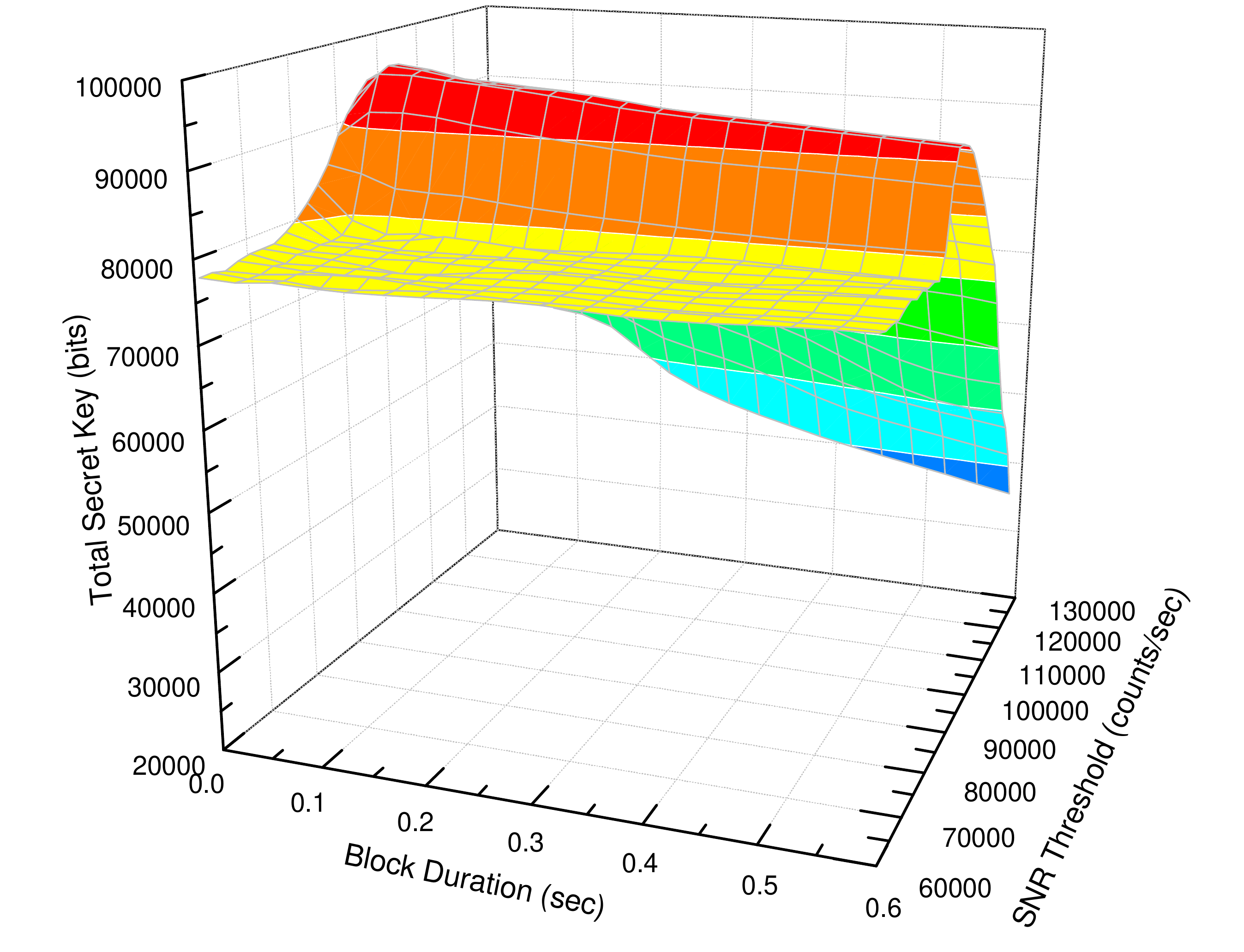}
    \caption{The total secret key for the high turbulence free-space experiment of Fig. \ref{fig.IQCPDTC} (b) plotted versus the block duration and the SNRF threshold, using a coincidence window of 5\unit{ns}. The optimum block duration was found to be 30\unit{ms}, while the optimum SNRF threshold was 95,000\unit{counts/sec} suitably applied on the timescale of the optimum block duration.}
    \label{fig.SecretKeyRates}
\end{figure}

After performing some initial simulations which showed the promise of the SNRF idea, we proceeded to implement the algorithm using the data gathered during the artificially increased turbulence experiment of Fig. \ref{fig.IQCPDTC} (b). There are three main parameters which affect the total secret key rate using the SNRF idea: the block duration, the SNRF threshold, and the coincidence window. The block duration refers to the time-scale on which the SNRF algorithm is applied and its optimum should be related to the time-scale of the atmospheric turbulence. The optimum SNRF threshold should be related to the mean background count rates observed during the experiment. Fig. \ref{fig.SecretKeyRates} shows the results of this analysis, with the total secret key generated from the 3\unit{min} block of data from Fig. \ref{fig.IQCPDTC} (b) plotted against the block duration and the SNRF threshold, for a coincidence window of 5\unit{ns}.

The key rates for the lower SNRF thresholds (closest to the front) in Fig. \ref{fig.SecretKeyRates} essentially show the secret key rate one would expect without implementing the SNRF algorithm (since little if any raw key is thrown away). As the SNRF threshold increases though (moving towards the top in Fig. \ref{fig.SecretKeyRates}), one can clearly see that the total secret key rate also increases until reaching a maximum at which point it quickly falls off since the SNRF cuts out too much raw key. Less obvious from the figure, but still important, there is a gradual improvement in the secret key rate as the block duration shrinks until a maximum is reached at which point the secret key rate gradually decreases once again. The optimum parameters for this data set were to use a block duration of 30\unit{ms} and a SNRF threshold of 95,000\unit{counts/sec} suitably applied on the timescale of the optimum block duration which increased the total secret key generated to 97,678\unit{bits} from the 78,009\unit{bits} generated when no SNRF was used. This represents an increase of 25.2\% in the total secret key generated from the \emph{same} raw key dataset.

As mentioned earlier, the secret key rate given by Eq. \ref{eq.InfiniteKeyRate} is improved due to two effects. First, the intrinsic error rate in the data is smaller causing the efficiency of the Cascade error correction algorithm \cite{BS94,SY00} used here to be improved from 1.2631 for the case of no SNRF to 1.2202 when a SNRF is used. This increased efficiency translates into fewer bits revealed during error correction and thus few bits sacrificed during privacy amplification. Secondly, the QBER measured during error correction is smaller, 4.30\% with a SNRF versus 5.51\% with none. This translates into less privacy amplification needed to ensure that the final secret key is secure against an eavesdropper.

\begin{table}[htbp]
  \centering
  \begin{tabular}{cccccc}
      Scenario & Raw key & Sifted key & Secret key & f & QBER\\
      \hline \hline
      No SNRF & 535,530 & 259,855 & 78,009 & 1.2697 & 5.51\% \\
      \hline
      Above SNRF & 466,441 & 226,279 & 97,678 & 1.2202 & 4.30\% \\
      \hline
      Below SNRF & 69,089 & 33,576 & - & - & 13.77\% \\
  \end{tabular}
  \caption{Measured values for: directly generating key, using the SNRF to generate key, and data discarded by the SNRF, for the high free-space turbulence experiment of Fig. \ref{fig.IQCPDTC} (b).}
  \label{tab.KeyGenValues}
\end{table}

In order to aid the potential security analysis of our SNRF idea, we also include a few other measured values pertinent to its implementation which are summarized in Tab. \ref{tab.KeyGenValues}. For the data set shown in Fig. \ref{fig.SecretKeyRates}, we kept 466,441 coincidences which made up our raw key while rejecting 69,089 coincidences generated from data blocks that were below the SNRF threshold. The size of the sifted key, where both Alice and Bob measured in the same basis, was 226,279 bits while 33,576 bits were rejected by the SNRF. As mentioned before, the QBER in this sifted key was 4.30\% while the QBER in the rejected data was 13.77\%. Here we can clearly see how utilizing the SNRF was able to increase our overall secret key rate by rejecting this small subset which turns out to have a much higher QBER.

While the preceding discussion nicely illustrated the usefulness of using the SNRF idea to produce a larger final key length from the same raw key rates, we now mention two ideas for how one would actually implement the SNRF algorithm in practice. The first idea would be in the case of a static scenario (fixed position free-space links) to use the first few minutes of an experiment to find the optimum SNRF threshold (since finding the optimum requires the full knowledge of the measurement results) which would then be used during the rest of the key exchange. For the case of a long distance key exchange with a quickly changing background one could periodically re-calculate the ideal SNRF threshold in order to continually operate with the optimum threshold. In the case of an exchange with an orbiting satellite, for example, one could periodically re-calculate the optimum every few seconds even as atmospheric conditions changed over the course of the key exchange. The beauty of the SNRF idea is that it can be applied completely offline, as the raw data from the key exchange can be saved and processed with the optimum threshold determined after the satellite has already passed overhead. Additionally, the user can apply as fine or coarse grained an analysis as they wish. A second possibility would be to have a catalogue of free-space parameter regimes and the corresponding optimum SNRF thresholds stored in a look-up table. Then one could continually monitor the free-space link statistics over the course of a key exchange (which require only the coincidence events to calculate) and pick the optimum threshold based on the measured free-space PDTC parameters.

Besides these implementation ideas, there are at least two other possibilities for future work to augment the protocol. The ideas are similar with the first being to use an adaptive block duration which expands and contracts depending on the single photon rates being observed. The optimum block duration of 30\unit{ms} found in this experiment was in a way a compromise between using larger blocks where fluctuations are averaged over and smaller blocks where there are unclear statistics. With an adaptive algorithm it would be possible to match the block duration more closely to the actual physical SNR variations during an experiment and thus increase the proportion of good transmission periods kept even more.

The second idea would seek to examine how the signal (singles rate) is correlated with the QBER (for instance, by plotting a 3D frequency (z-axis) histogram of signal (x-axis) versus QBER (y-axis)). With this correlation plot, one could try to predict what the most likely QBER would be for a given signal level. Then one could apply a finer filtering scheme, for instance, grouping data blocks into the three classes: low QBER, medium QBER ($<11\%$), and high QBER ($>11\%$). Certainly the high QBER blocks should be discarded because they actually cost key. But while the medium QBER blocks may still have a QBER higher than that of the intrinsic system due to background light, they would still contribute positively to the key. Processing them separately from the low QBER blocks however would allow one to optimize the algorithms used for each subset to make them as efficient as possible.

\begin{figure}[htp]
    \centering
    \includegraphics[width=0.9\textwidth]{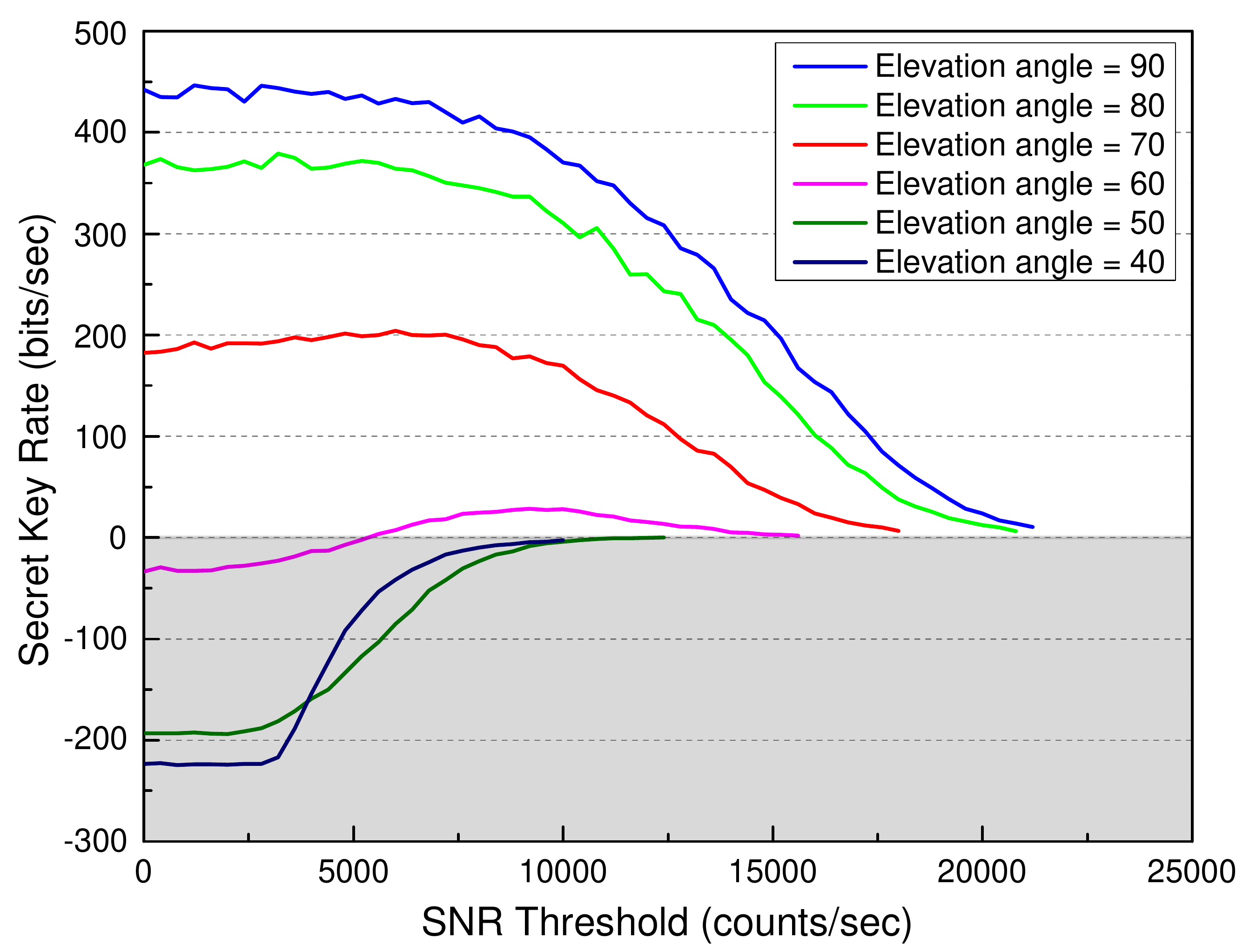}
    \caption{The simulated secret key rates for a LEO QKD satellite for various elevation angles and the expected number of background counts and free-space PDTC \cite{BMHHEHKHDGLLJ12}. We assume the entangled source on the satellite operates at 100\unit{MHz} with an intrinsic QBER of 2.5\%. Negative rates in the shaded region below the 0 axis would produce no secret key and are shown to emphasize the fact that utilizing the SNRF can produce a positive key rate where otherwise none would be possible.}
    \label{fig.KeyRateVsBackgroundSim}
\end{figure}

The full power of the SNRF idea is realized in cases with a high background where the accidental coincidence rate approaches the same order of magnitude as the QKD signal. Recent work for the case of performing QKD with an orbiting satellite \cite{BMHHEHKHDGLLJ12} has shown that one will indeed be operating in a high background regime where the SNRF idea will prove very useful. Further to this point, depending on the level of the background noise, the simulations in Fig. \ref{fig.KeyRateVsBackgroundSim} show that the SNRF idea can be used to produce a secret key when the background would otherwise have prevented it. Fig. \ref{fig.KeyRateVsBackgroundSim} plots the secret key rates for exchanging key with with an orbiting LEO QKD satellite carrying an entangled photon source with a pair production rate of 100\unit{MHz} and an intrinsic QBER of 2.5\% for various elevation angles and the expected number of background counts and free-space PDTC \cite{BMHHEHKHDGLLJ12}. These initial results show that the SNRF idea would allow us to generate secret key from many more satellite passes occurring at elevation angles of $70^{\circ}$ or less; though a more detailed analysis would have to take into account the statistics of satellite passes over a year which must integrate over the various elevation angles. Nonetheless, the SNRF idea should prove particularly useful as the most probable passes for a LEO satellite occur at elevation angles much less than $90^{\circ}$ which otherwise would render them useless due to the high free-space link fluctuations, high background, and low SNR experienced. Thus, we are very confident that the SNRF idea will prove extremely useful in high background situations such as in satellite QKD, long distance terrestrial free-space links, or daylight QKD experiments.

\section{Conclusions}

In conclusion, we have used an entanglement based free-space QKD system to study the link statistics generated during the fluctuating free-space transmission of entangled photon pairs. Simulating a free-space channel with a high amount of turbulence allowed us to recover the theoretical prediction of a log-normal distribution for the statistics of the transmission coefficient. Using insights from this analysis, we studied the effect of link fluctuations on free-space decoy state QKD and found that the static channel approximation typically assumed is valid for the regimes where such systems are typically operated. Lastly, we studied the implementation of a signal-to-noise ratio filter in order to increase the overall secret key rate by rejecting measurements during periods of low transmission efficiency which tend to have a larger QBER due to a higher proportion of background events to actual entangled pair detection in the raw key. Using this SNRF, we were able to increase the final secret key rate by 25.2\% using the \emph{same} raw signals for a particular experimental run. Further, we showed simulations that indicate that the SNRF idea will be extremely useful in terrestrial long distance free-space experiments and experiments exchanging a key with a LEO satellite allowing one to generate a secret key from many passes that would otherwise have been useless.


\ack
Support for this work by NSERC, CSA, CIFAR, CFI, ORF, OCE, ERA, and the Bell family fund is gratefully acknowledged. One of us (BH) would like to thank the EU-Canada Programme for Cooperation in Higher Education, Training, and Youth which sponsored her exchange with the IQC during this project. The authors would like to thank M. Toyoshima, B. Higgins, and N. L\"utkenhaus for helpful discussions about free-space satellite communication, turbulence induced link fluctuations, and the quantum security of such schemes.

During the final preparation of this manuscript the authors became aware of similar work \cite{CTDGUVV12} which examined the impact of turbulence on long range quantum and classical channels, and put forward a similar SNR proposal based on measuring a secondary beacon laser. While the work measuring the PDTC of the free-space links is similar, we stress that we have already implemented such an SNRF \emph{directly} on the detected signal and illustrated the improvements possible.

Lastly, the authors would like to thank the anonymous referees for their many comments, which were very useful in improving the quality of this paper.


\section*{References}
\bibliographystyle{unsrt}	
\bibliography{Paper9_Bibliography}

\begin{thebibliography}{10}

\bibitem{SFIK11}
M.~Sasaki, M.~Fujiwara, H.~Ishizuka, W.~Klaus, K.~Wakui, M.~Takeoka, S.~Miki,
  T.~Yamashita, Z.~Wang, A.~Tanaka, K.~Yoshino, Y.~Nambu, S.~Takahashi,
  A.~Tajima, A.~Tomita, T.~Domeki, T.~Hasegawa, Y.~Sakai, H.~Kobayashi,
  T.~Asai, K.~Shimizu, T.~Tokura, T.~Tsurumaru, M.~Matsui, T.~Honjo, K.~Tamaki,
  H.~Takesue, Y.~Tokura, J.~F. Dynes, A.~R. Dixon, A.~W. Sharpe, Z.~L. Yuan,
  A.J. Shields, S.~Uchikoga, M.~{Legr\'e}, S.~Robyr, P.~Trinkler, L.~Monat,
  J.B. Page, G.~Ribordy, A.~Poppe, A.~Allacher, O.~Maurhart, T.~{L\"anger},
  M.~Peev, and A.~Zeilinger.
\newblock Field test of quantum key distribution in the tokyo qkd network.
\newblock {\em Opt. Exp.}, 19:10387--10409, 2011.

\bibitem{PPAB09}
M.~Peev, C.~Pacher, R~{All\'eaume}, C.~Barreiro, J.~Bouda, W.~Boxleitner,
  T~Debuisschert, E.~Diamanti, M.~Dianati, J.F. Dynes, S.~Fasel, S.~Fossier,
  M.~{F\"urst}, J.D. Gautier, O.~Gay, N.~Gisin, P.~Grangier, A.~Happe,
  Y.~Hasani, M.~Hentschel, H.~{H\"ubel}, G.~Humer, T.~{L\"anger}, M.~{Legr\'e},
  R.~Lieger, J.~Lodewyck, T.~{Lor\"unser}, N.~{L\"utkenhaus}, A.~Marhold,
  T.~Matyus, O.~Maurhart, L.~Monat, S.~Nauerth, J.B. Page, A.~Poppe,
  E.~Querasser, G.~Ribordy, S.~Robyr, L.~Salvail, A.W. Sharpe, A.J. Shields,
  D.~Stucki, M.~Suda~C. Tamas, T.~Themel, T.~Thew, Y.~Thoma, A.~Treiber,
  P.~Trinkler, R.~Tualle-Brouri, F.~Vannel, N.~Walenta, H.~Weier,
  H.~Weinfurter, I.~Wimberger, Z.L. Yuan, H.~Zbinden, and A.~Zeilinger.
\newblock The secoqc quantum key distribution network in vienna.
\newblock {\em New J. Phys.}, 11:075001, 2009.

\bibitem{SBCDLP09}
V.~Scarani, H.~Bechmann-Pasquinucci, N.J. Cerf, M.~{Du\u sek},
  N.~{L\"utkenhaus}, and M.~Peev.
\newblock A framework for practical quantum cryptography.
\newblock {\em Rev. Mod. Phys.}, 81:1301--1350, 2009.

\bibitem{GRTZ02}
N.~Gisin, G.~Ribordy, W.~Tittel, and H.~Zbinden.
\newblock Quantum cryptography.
\newblock {\em Rev. Mod. Phys.}, 74:145--195, 2002.

\bibitem{JRYYZXWYHJYYCPP10}
X.M. Jin, J.G. Ren, B.~Yang, Z.H. Yi, F.~Zhou, X.F. Xu, S.K. Wang, D.~Yang,
  Y.F. Hu, S.~Jiang, T.~Yang, H.~Yin, K.~Chen, C.Z. Peng, and J.W. Pan.
\newblock Experimental free-space quantum teleportation.
\newblock {\em Nature Photonics}, 4:376--381, 2010.

\bibitem{EBHWSL09}
D.~Elser, T.~Bartley, B.~Heim, Ch. Wittmann, D.~Sych, and G.~Leuchs.
\newblock Feasibility of free space quantum key distribution with coherent
  polarization states.
\newblock {\em New J. Phys.}, 11:045014, 2009.

\bibitem{ECLW08}
C.~Erven, C.~Couteau, R.~Laflamme, and G.~Weihs.
\newblock Entangled quantum key distribution over two free-space optical links.
\newblock {\em Opt. Exp.}, 16:16840--16853, 2008.

\bibitem{PFHLK08}
M.P. Peloso, I.~Gerhardt, C.~Ho, A.~Lamas-Linares, and C.~Kurtsiefer.
\newblock Daylight operation of a free space, entanglement-based quantum key
  distribution system.
\newblock {\em New J. Phys.}, 11:045007, 2009.

\bibitem{UTSWSLBJPTOFMRSBWZ07}
R.~Ursin, F.~Tiefenbacher, T.~Schmitt-Manderbach, H.~Weier, T.~Scheidl,
  M.~Lindenthal, B.~Blauensteiner, T.~Jennewein, J.~Perdigues, P.~Trojek,
  B.~{\"Omer}, M.~{F\"urst}, M.~Meyenburg, J.G. Rarity, Z.~Sodnik, C.~Barbieri,
  H.~Weinfurter, and A.~Zeilinger.
\newblock Entanglement-based quantum communication over 144km.
\newblock {\em Nature Physics}, 3:481--486, 2007.

\bibitem{NHMPW02}
J.E. Nordholt, R.J. Hughes, G.L. Morgan, C.G. Peterson, and C.C. Wipf.
\newblock Present and future free-space quantum key distribution.
\newblock {\em Proc. SPIE}, 4635:116, 2002.

\bibitem{MagiQ}
{MagiQ Technologies}.
\newblock {MagiQ Technologies}, 2008.
\newblock http://www.magiqtech.com/.

\bibitem{idQuantique}
{idQuantique}.
\newblock idquantique, 2011.
\newblock http://www.idquantique.com/.

\bibitem{QuintessenceLabs}
{Quintessence Labs}.
\newblock Quintessence labs, 2011.
\newblock http://www.quintessencelabs.com/.

\bibitem{SeQureNet}
{SeQureNet}.
\newblock Sequrenet, 2011.
\newblock http://www.sequrenet.fr/.

\bibitem{DYDSS08}
A.R. Dixon, Z.L. Yuan, J.F. Dynes, A.W. Sharpe, and A.J. Shields.
\newblock Gigahertz decoy quantum key distribution with 1mbit/s secure key
  rate.
\newblock {\em Opt. Exp.}, 16:18790--18797, 2008.

\bibitem{RTGK02}
J.G. Rarity, P.R. Tapster, P.M. Gorman, and P.~Knight.
\newblock Ground to satellite secure key exchange using quantum cryptography.
\newblock {\em New J. Phys.}, 4:82, 2002.

\bibitem{AJPLZ03}
M.~Aspelmeyer, T.~Jennewein, M.~Pfennigbauer, W.~Leeb, and A.~Zeilinger.
\newblock Long-distance quantum communication with entangled photons using
  satellites.
\newblock {\em IEEE J. of Selected Topics in Quantum Electronics}, 9:1541,
  2003.

\bibitem{AFMMCPAJUSBRLBWZ07}
J.~Armengol, B.~Furch, C.~de~Matos, O.~Minster, L.~Cacciapuoti,
  M.~Pfennigbauer, M.~Aspelmeier, T.~Jennewein, R.~Ursin,
  T.~Schmitt-Manderbach, G.~Baister, J.~Rarity, W.~Leeb, C.~Barbieri,
  H.~Weinfurter, and A.~Zeilinger.
\newblock Quantum communications at esa - towards a space experiment on the
  iss.
\newblock {\em Acta Astronautica}, 63:165--178, 2008.

\bibitem{BMHHEHKHDGLLJ12}
J.P. Bourgoin, E.~Meyer-Scott, B.L. Higgins, B.~Helou, C.~Erven, H.~Huebel,
  B.~Kumar, D.~Hudson, I.~D'Souza, R.~Girard, N.~{L\"utkenhaus}, R.~Laflamme,
  and T.~Jennewein.
\newblock A comprehensive design and performance analysis of leo satellite
  quantum communications.
\newblock {\em in preparation}, 2012.

\bibitem{Kol41}
A.N. Kolmogorov.
\newblock The local structure of turbulence in incompressible viscous fluid for
  very large reynolds numbers.
\newblock {\em Doklady ANSSSR}, 30:301--304, 1941.

\bibitem{Tat71}
V.I. Tatarskii.
\newblock US Department of Commerce, Springfield, VA, 1971.

\bibitem{Fan75}
R.L. Fante.
\newblock Electromagnetic beam propagation in turbulent media.
\newblock {\em Proceedings of the {IEEE}}, 63:1669--1692, 1975.

\bibitem{Fan80}
R.L. Fante.
\newblock Electromagnetic beam propagation in turbulent media: An update.
\newblock {\em Proceedings of the {IEEE}}, 68(11):1424-- 1443, 1980.

\bibitem{AP05}
L.C. Andrews and R.L. Phillips.
\newblock SPIE Press, 2005.

\bibitem{APY95}
L.C. Andrews, R.L. Phillips, and P.T. Yu.
\newblock Optical scintillations and fade statistics for a
  satellite-communication system.
\newblock {\em Appl. Opt.}, 34(33):7742--7751, 1995.

\bibitem{Fri67}
D.L. Fried.
\newblock Scintillation of a {Ground-to-Space} laser illuminator.
\newblock {\em Journal of the Optical Society of America}, 57(8):980--983,
  1967.

\bibitem{Sha11}
J.H. Shapiro.
\newblock Scintillation has minimal impact on far-field bennett-brassard 1984
  protocol quantum key distribution.
\newblock {\em Phys. Rev. A}, 84:032340, 2011.

\bibitem{MCPH04}
P.W. Milonni, J.H. Carter, C.G. Peterson, and R.J. Hughes.
\newblock Effects of propagation through atmospheric turbulence on photon
  statistics.
\newblock {\em J. Opt. B: Quantum Semiclass. Opt.}, 6:S742--S745, 2004.

\bibitem{VSV12}
D.Y. Vasylyev, A.A. Semenov, and W.~Vogel.
\newblock Towards global quantum communication: Beam wandering preserves
  quantumness.
\newblock {\em Phys. Rev. Lett.}, 108:220501, 2012.

\bibitem{SV10}
A.A. Semenov and W.~Vogel.
\newblock Entanglement transfer through the turbulent atmosphere.
\newblock {\em Phys. Rev. A}, 81:023835, 2010.

\bibitem{SV09}
A.A. Semenov and W.~Vogel.
\newblock Quantum light in the turbulent atmosphere.
\newblock {\em Phys. Rev. A}, 80:021802, 2009.

\bibitem{Smi93}
F.G. Smith, editor.
\newblock {\em Atmospheric Propagation of Radiation}, volume~2.
\newblock SPIE Optical Engineering Press, Bellingham, WA, USA, 1993.

\bibitem{EHRLW10}
C.~Erven, D.~Hamel, K.~Resch, R.~Laflamme, and G.~Weihs.
\newblock Entanglement based quantum key distribution using a bright sagnac
  entangled photon source.
\newblock In O.~Akan, P.~Bellavista, J.~Cao, F.~Dressler, D.~Ferrari M., Gerla,
  H.~Kobayashi, S.~Palazzo, S.~Sahni, X.~Shen, M.~Stan, J.~Xiaohua, A.~Zomaya,
  G.~Coulson, A.~Sergienko, S.~Pascazio, and P.~Villoresi, editors, {\em
  Quantum Communication and Quantum Networking}, volume~36 of {\em Lecture
  Notes of the Institute for Computer Sciences, Social Informatics and
  Telecommunications Engineering}, pages 108--116. Springer Berlin Heidelberg,
  2010.

\bibitem{KFW06}
T.~Kim, M.~Fiorentino, and F.N.C. Wong.
\newblock Phase-stable source of polarization-entangled photons using a
  polarization sagnac interferometer.
\newblock {\em Phys. Rev. A}, 73:012316, 2006.

\bibitem{FHPJZ07}
A.~Fedrizzi, T.~Herbst, A.~Poppe, T.~Jennewein, and A.~Zeilinger.
\newblock A wavelength-tunable fiber-coupled source of narrowband entangled
  photons.
\newblock {\em Opt. Exp.}, 15:15377, 2007.

\bibitem{BGN00}
G.~Brida, M.~Genovese, and C.~Novero.
\newblock An application of two-photon entangled states to quantum metrology.
\newblock {\em J. Mod. Opt.}, 47:2099--2104, 2000.

\bibitem{GLLP04}
D.~Gottesman, H.-K. Lo, N.~{L\"utkenhaus}, and J.~Preskill.
\newblock Security of quantum key distribution with imperfect devices.
\newblock {\em Quant. Info. Compu.}, 4:325--360, 2004.

\bibitem{Hwa03}
W.Y. Hwang.
\newblock Quantum key distribution with high loss: Toward global secure
  communication.
\newblock {\em Phys. Rev. Lett.}, 91:057901, 2003.

\bibitem{LMC05}
H.K. Lo, X.~Ma, and K.~Chen.
\newblock Decoy state quantum key distribution.
\newblock {\em Phys. Rev. Lett.}, 94:230504, 2005.

\bibitem{MQZL05}
X.~Ma, B.~Qi, Y.~Zhao, and H.K. Lo.
\newblock Practical decoy state for quantum key distribution.
\newblock {\em Phys. Rev. A}, 72:012326, 2005.

\bibitem{HMDFLLA06}
J.~Heersink, C.~Marquardt, R.~Dong, R.~Filip, S.~Lorenz, G.~Leuchs, and U.L.
  Andersen.
\newblock Distillation of squeezing from non-gaussian quantum states.
\newblock {\em Phys. Rev. Lett.}, 96:253601, 2006.

\bibitem{DLHMFLA08}
R.~Dong, M.~Lassen, J.~Heersink, C.~Marquardt, R.~Filip, G.~Leuchs, and U.L.
  Andersen.
\newblock Experimental entanglement distillation of mesoscopic quantum states.
\newblock {\em Nature Physics}, 4:919--923, 2008.

\bibitem{MFL07b}
X.~Ma, C.H. Fung, and H.K. Lo.
\newblock Quantum key distribution with entangled photon sources.
\newblock {\em Phys. Rev. A}, 76:012307, 2007.

\bibitem{BS94}
G.~Brassard and L.~Salvail.
\newblock Secret-key reconciliation by public discussion.
\newblock {\em Lect. Notes Comput. Sci.}, 765:410, 1994.

\bibitem{SY00}
T.~Sugimoto and K.~Yamazaki.
\newblock A study on secret key reconciliation protocol "cascade".
\newblock {\em IEICE Trans. Fundamentals}, E83A No. 10:1987, 2000.

\bibitem{Gal62}
R.G. Gallager.
\newblock Low density parity check codes.
\newblock {\em IRE Trans. Info. Theory}, IT-8:21--28, 1962.

\bibitem{MN97}
D.J.C. MacKay and R.M. Neal.
\newblock Near shannon limit performance of low density parity check codes.
\newblock {\em Electronics Letters}, 33:457--458, 1997.

\bibitem{Pea04}
D.~Pearson.
\newblock High-speed qkd reconciliation using forward error correction.
\newblock In {\em QCMC}, 2004.

\bibitem{MCMHT09}
I.L. Martinez, P.~Chan, X.~Mo, S.~Hosier, and W.~Tittel.
\newblock Proof-of-concept of real-world quantum key distribution with quantum
  frames.
\newblock {\em New J. Phys.}, 11:095001, 2009.

\bibitem{CW79}
J.L. Carter and M.N. Wegman.
\newblock Universal classes of hash functions.
\newblock {\em Journal of Computer and System Sciences}, 18:143, 1979.

\bibitem{Kra94}
H.~Krawczyk.
\newblock Lfsr-based hashing and authentication.
\newblock {\em Advances in Cryptology - CRYPTO '94 - Lecture Notes in Computer
  Science}, 839:129--139, 1994.

\bibitem{CCHGKMTONMSWFRGGHTB11}
P.J. Clarke, R.J. Collins, P.A. Hiskett, M.J. Garcia-Martinez, N.J. Krichel,
  A.~McCarthy, M.G. Tanner, J.A. O'Connor, C.M. Natarajan, S.M. Miki,
  M.~Sasaki, Z.~Wang, M.~Fujiwara, I.~Rech, M.~Ghioni, A.~Gulinatti,
  R.~Hadfield, P.D. Townsend, and G.S. Buller.
\newblock Analysis of detector performance in a gigahertz clock rate quantum
  key distribution system.
\newblock {\em New J. Phys.}, 13:075008, 2011.

\bibitem{UHPWMLF12}
V.C. Usenko, B.~Heim, C.~Peuntinger, C.~Wittmann, C.~Marguardt, G.~Leuchs, and
  R.~Filip.
\newblock Entanglement of gaussian states and the applicability to quantum key
  distribution over fading channels.
\newblock {\em eprint}, quant-ph/1208.4307, 2012.

\bibitem{CTDGUVV12}
I.~Capraro, A.~Tomaello, A.~Dall\'Arche, F.~Gerlin, R.~Ursin, G.~Vallone, and
  P.~Villoresi.
\newblock Impact of turbulence in long range quantum and classical
  communications.
\newblock {\em eprint}, quant-ph/1207.6931, 2012.

\end{thebibliography}

\end{document}